\documentclass[apj]{emulateapj}

\newcommand{\apjvec}[1]{\mbox{\boldmath{$#1$}}}

\newcommand{\vcn}{\apjvec{n}}
\newcommand{\vcr}{\apjvec{r}}

\begin{document}

\title{Searchable Sky Coverage of Astronomical Observations: Footprints and Exposures}
\journalinfo{To be submitted to \pasp}

\author{Tam\'as Budav\'ari, Alexander S. Szalay and Gy\"orgy Fekete}
\affil{Dept.\ of Physics and Astronomy, The Johns Hopkins University, 3400 North Charles Street, Baltimore, MD 21218}


\shortauthors{Budav\'ari, Szalay \& Fekete}
\shorttitle{Searchable Sky Coverage}

\begin{abstract}
Sky coverage is one of the most important pieces of information about
astronomical observations. We discuss possible representations, and
present algorithms to create and manipulate shapes consisting of
generalized spherical polygons with arbitrary complexity and size on
the celestial sphere.
This shape specification integrates well with our Hierarchical Triangular
Mesh indexing toolbox, whose performance and capabilities are enhanced
by the advanced features presented here.
Our portable implementation of the relevant spherical geometry
routines comes with wrapper functions for database queries, which
are currently being used within several scientific catalog archives
including the Sloan Digital Sky Survey, the Galaxy Evolution Explorer
and the Hubble Legacy Archive projects as well as the Footprint Service
of the Virtual Observatory.
\end{abstract}


\section{Introduction} \label{sec:intro}

Astronomers have to keep accurate records of where their observations are
located on the sky.
Beyond the direction and angular size of the field of view, we have
detailed information available about the precise sky coverage derived 
from the exact shape of our detectors.
This coverage is invaluable for most statistical studies, e.g.,
luminosity functions or especially analyses of spatial clustering.

The de facto standard of the Flexible Image Transport System 
\citep[FITS;][]{fits} has reserved keywords for the World Coordinate 
System \citep[WCS;][]{wcs} specification, and parameters that specify 
the transformation from image pixels to sky coordinates (and reverse),
are present in the header of most FITS files.
While the WCS is perfectly adequate for individual exposures, multiple 
observations are difficult to describe in a single system. Every field 
has potentially a separate coordinate system, hence moving from field to 
field is convoluted, and makes it cumbersome to answer even simple 
questions, e.g., whether two separate fields overlap.

The footprint of a large-area survey might be complicated but the
small-scale irregularities are even more problematic. Not only the depth of
a survey changes as a function of the position on the sky, but parts of the
observations are often censored for various reasons, such as bright stars
blocking the view, satellite trails, artifacts from reflections.
If we would like to represent all these on the sky, we need a scalable 
solution that works for shapes of arbitrary complexity and size from the 
subpixel level to the entire sky. We need tools to create and manipulate 
these descriptions right there where the data are and utilize the information
efficiently.

Geographic Information Systems (GIS) were designed with a similar goal in mind.
There are however subtle differences, that are large enough that GIS systems
are not quite applicable to astronomy directly. 
The modern mapping systems do not extend much beyond the basic features of 
projected maps but provide very efficient tools for finding nearby places and
shortest routes, etc.
Even the most complicated GIS shapes are limited to spherical
polygons whose sides are great circles (or straight lines in the projection.)
In astronomy, the approximations with such polygons would be unacceptably inaccurate
or prohibitively redundant, hence there is need for an extended set of features to
represent the geometries of the observations and surveys.

Some of the concepts discussed in the paper are not new and
have been introduced and studied previously in different fields.
Books by \citet{samet89,samet90} detail the quad-trees for spatial
searches that \citet{barrett94} advocated using for astronomy data.
\cite{fekete90a,fekete90b,short95} developed an icosahedron-based methodology for
Earth sciences, and \citet{kunszt00, kunszt01} introduced the convex description and the Hierarchical Triangular Mesh,
a triangulation that was also used by \citet{lee98} with a different numbering scheme.
A similar representation of shapes is also found in \citet{mangle1} and \citet{mangle2}.
\citet{goodchild91,goodchild92} and \citet{song00} introduced the Discrete
Global Grid for GIS systems, an equal area variant of the same triangulation idea.
The integration to relational databases is discussed in \citet{gray04}.

The goal this paper is to provide the astronomy community with a complete
and consistent view of the current and much improved methodology built on a new
fully functional spherical geometry framework, and describe the implementations 
and interfaces used in several astronomy archives and tools today including 
the Sloan Digital Sky Survey \citep{cas}, the Hubble Legacy Archive \citep{greene07},
the Galaxy Evolution Explorer and the NVO Footprint Service \citep{adass_footprint}.
In Section~\ref{sec:shapes} we describe how to specify spherical shapes and manipulate them.
Section~\ref{sec:geom} deals with the spherical geometry of the region representations
and
Section~\ref{sec:htm} discusses an efficient search method
based on the Hierarchical Triangular Mesh.
In Section~\ref{sec:lib} we provide details of our
software solution including the implementation of
the database routines and their usage.
Section~\ref{sec:sum} summarizes the main results, along with
current and future applications in astronomy.

\section{Shapes on the Celestial Sphere} \label{sec:shapes}

In cartography, maps are typically local projections, 
e.g., the pages of an atlas that just overlap so that they provide
enough reference for navigating a road or a trail,
and in general, moving from one projection to another can be quite difficult.
%
%
In astronomy, the usage pattern is different and
usually more global that warrants the use of a true spherical geometry.

Spherical polygons are closed geometric figures over the sphere, formed by
arcs of great circles. Generalized spherical polygons (GSPs) are similar, 
but their arcs can also be small circles. These are conceptually simple and yet 
versatile enough to represent most common shapes in astronomy.
They easily describe circles, rectangles (even with curved sides), and
where they cannot precisely track a boundary, an accurate approximation
is possible by a short series of arc segments. Not only do they conveniently 
describe spherical shapes but they also provide a very compact representation.
For example, the vertices and the equations of the edges (arcs) define the
{\em outline} of a generalized spherical polygon, and its inside can be
determined by the order of the vertices. One convention is to define the 
inside to be to the left as one traverses around the polygon.
Either a small or a great circle can be defined as the intersection
of the unit sphere with a 3-D plane. This enables us to use another, dual
representation, by using half-space contraints that define the interior 
{\it surface} area of these spherical circles, instead of their outline. 
If we embed the surface of the unit sphere in a 3-dimensional Eucledian space, 
we can use 3D directed planar (halfspace) constraints and their boolean 
combinations to select parts of the sphere that represent various spherical 
shapes of this family describable by GSPs.
Regions that cannot be represented this way would be the curves defined
by intersections of higher order surfaces with the sphere, like quadrics.
These two alternative descriptions are formally equivalent but have different
properties that make them preferable for certain types of problems.
Both have advantages and disadvantages but we do not have to choose one.
In this section, we first elaborate on the surface or convex representation
in detail with special emphasis on its strengths, and point out how to
obtain the outline algorithmically.

\subsection{Halfspaces, Convexes and Regions}

Going from a two-dimensional representation of spherical shapes
to a three-dimensional description provides a uniform framework
with readily available geometrical concepts to build on.
A {\em halfspace} is a directed plane that splits the 3-D space in two.
In our context, it represents a spherical cap when intersected
with the unit sphere.
It is defined by a direction, i.e., a unit vector $\vcn$, and a signed scalar 
offset $c$, measured along the normal vector from the origin. Using a unit-sphere 
centered on the origin of a Cartesian coordinate system, the parameter $c$ 
can take values between 1 (an infinitely small cap) and -1 (the whole sky). The 
value of 0 corresponds to a special case and represents half the sky cut along
a great circle. In general, $c$ is the cosine of the angular radius of the cap.
The caps with a negative $c$ are called {\it holes}.
Another common shape in astronomy is a rectangle in the tangent plane defined 
by the geometry of the detector. Since a straight line in any tangent plane 
projects onto a great circle on the surface of the sphere, the rectangle is a
convex formed by the intersection of four zero-offset halfspaces, whose normal 
vectors point in the direction of the neighboring vertices' cross products. 

An intersection of halfspace constraints represents a (possibly open) 3-D {\em 
convex} polyhedron, which in turn describes a more complicated shape when 
it intersects the unit sphere.
Despite their simplicity in 3-D, convexes can define a large variety of
spherical shapes, e.g., rings, diamonds or other polygons of ``straight''
or curved sides. In fact, they can even represent multiply connected shapes of 
arbitrarily large topological complexity. For example, the vertices and edges of 
a large enough cube centered on the sphere can define eight disjoint 
generalized spherical triangles as they poke through the surface.

Nevertheless, convexes are constrained in 3-D and, hence are limited in what
types of spherical shapes can describe.
More general spherical {\em regions} can be defined as the unions of convexes.
This hierarchy of halfspaces, convexes and regions and their negation (or 
difference) provides a complete algebra over these spherical regions, enabling 
extreme flexibility and efficiency.

\subsection{Point in a Region}

One of the most common tasks is the containment test to decide whether a
point is inside a region. We start with the most basic elements of the
region, the halfspaces. A point (unit vector $\vcr$) is inside a 
halfspace ($\vcn,c$), if the dot product of the two vectors is greater 
than the offset,
\begin{equation}
\vcn \cdot \vcr > c.
\end{equation}
While this is straightforward to see, its computational simplicity is 
striking, and has serious consequences for the performance of
any implementation using this formalism.
For example, the computation requires only three multiplications and one
comparison for a circle and four times that for a spherical quadrangle,
regardless of the size and actual shape.

Testing a point against a convex simply consists of checking whether 
the point is inside all its halfspaces. If yes, the point is inside; 
otherwise outside.
As a region is the union of its convexes, it contains a point
if any of its convexes contains the point.

\subsection{Boolean Algebra of Regions} \label{sec:alg}

Within this framework, the boolean algebra of regions maps very well
onto basic set operations on the collections of halfspaces and convexes.
The set of regions is closed for the operations one routinely performs 
when deriving survey specific geometries. The same is not true for the convexes.

The {\em union} of two or more regions is a region that includes all the convexes,
\begin{eqnarray}
R^{(1)} \cup R^{(2)} & = & C^{(1)}_1 \cup \dots \cup C^{(1)}_n \cup C^{(2)}_1 \cup \dots \cup C^{(2)}_m \\
 & = &  C_1 \cup \dots \cup C_{n+m} = R
\end{eqnarray}
by definition.
Also the {\em intersection} of two or more convexes is a convex that includes all their halfspaces,
\begin{eqnarray}
C^{(1)} \cap C^{(2)} & = & H^{(1)}_1 \cap \dots \cap H^{(1)}_n \cap H^{(2)}_1 \cap \dots \cap H^{(2)}_m \\
 & = &  H_1 \cap \dots \cap H_{n+m} = C
\end{eqnarray}
and, in turn, the intersection of regions is a region whose convexes are
the pairwise intersections of the convexes, e.g., for two
\begin{equation}
R^{(1)} \cap R^{(2)} = \bigcup_{i,j} \left( C^{(1)}_i \cap C^{(2)}_j \right)
\end{equation}
It is straightforward to define the {\em differences} of halfspaces, convexes,
and regions, even if not as simple as the above operations.
Let us first look at two halfspaces. We see that their difference is
\begin{equation}
H_1 \setminus H_2 = H_1 \cap \bar{H}_2
\end{equation}
where $\bar{H}$ is the {\em negate} or inverse of $H$, which
is obtained by simply flipping the sign of its normal vector and the offset.
Similarly, subtracting a halfspace from a convex is also simple as is
the subtraction from a region.
It might look logical to express the difference of two convexes as a
region whose convexes are
\begin{equation}
C^{(1)} \setminus C^{(2)} = \bigcup_{i} \left( C^{(1)} \setminus H^{(2)}_i \right)
    = \bigcup_i \left( C^{(1)} \cap \bar{H}^{(2)}_i \right)
\end{equation}
but the constructed convexes would overlap, hence the following procedure
is preferred to avoid the overlaps
\begin{equation}
C^{(1)} \setminus C^{(2)} = \bigcup_{i} \left( C^{(1)} \cap \bar{H}^{(2)}_i 
	\bigcap_k^{i-1} H^{(2)}_k \right)
\end{equation}
Also we difference a region and a convex by subtracting the convex from
all the convexes of the region.
By substituting the all-sky coverage into $C^1$ of the previous equation,
we get the region of a negated convex
\begin{equation}
\bar{C} = \bigcup_i^n \left( \bar{H}_i \bigcap_{k=1}^{i-1} H_k \right)
\end{equation}
The negate of a region is in turn the intersection of its inverted convexes,
by a simple application of deMorgan's rules.
From these formulas we can also see that it is not enough to stop at
the level of the convexes, because even if the basic building blocks of
a particular geometry are simple, subsequent operations quickly yield
more complicated descriptions that can only be represented as a region.

\begin{figure}
\epsscale{0.7}
\plotone{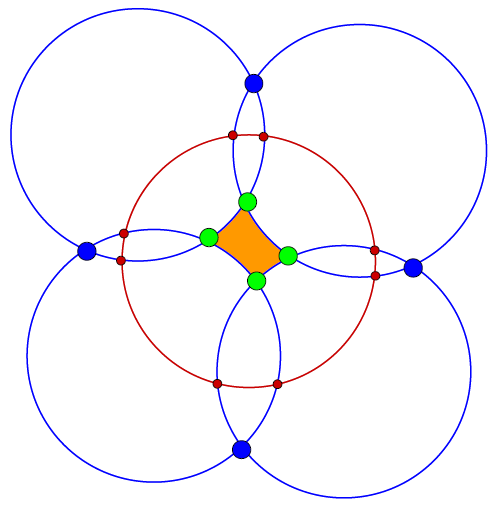}
\caption{The concave diamond shape is described by four halfspaces of
negative offsets, whose arcs draw its outline, plus a fifth constraint
that separates the diamond from the remainder of the sphere outside the
four holes.}
\label{fig:diamond}
\end{figure}

\begin{figure*}
\epsscale{0.95}
\plotone{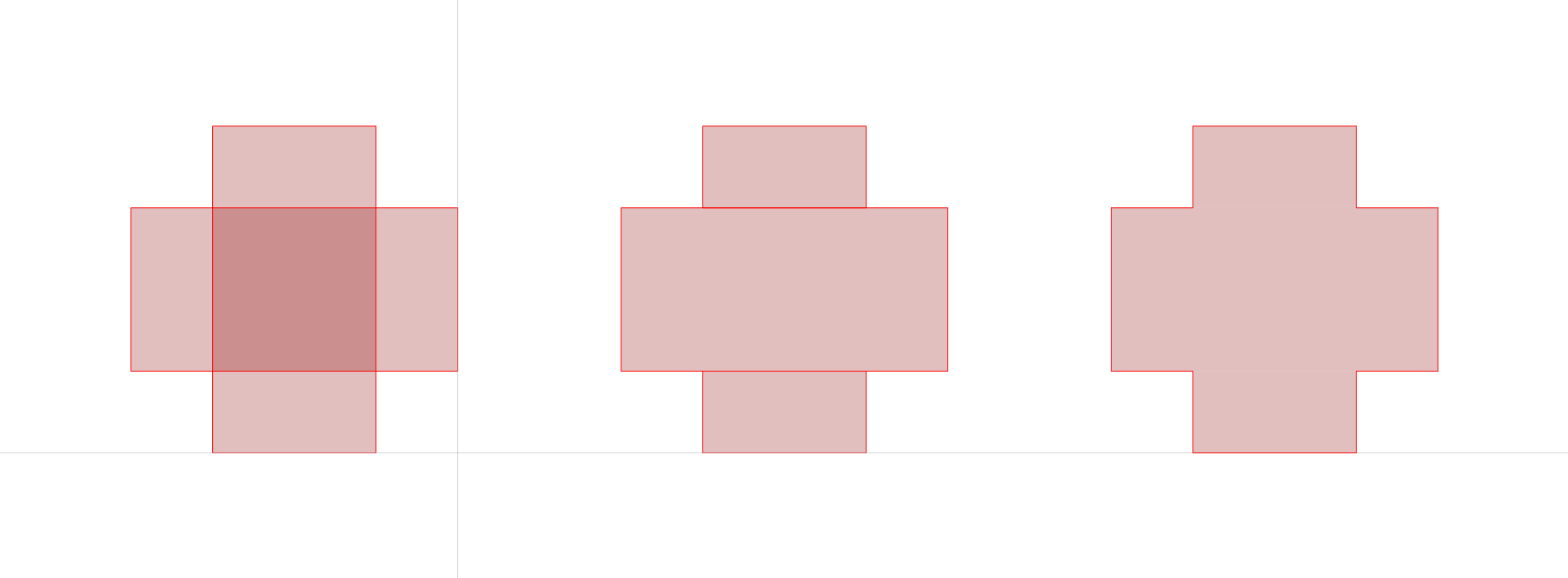}
\caption{The union of two overlapping rectangles ({\em left})
is turned into three disjoint convexes ({\em middle}) in the process
of simplification. The outline ({\em right}) eliminates those parts of the arcs
that are internal and cancel out.}
\label{fig:union}
\end{figure*}

\begin{figure*}
\epsscale{0.95}
\plotone{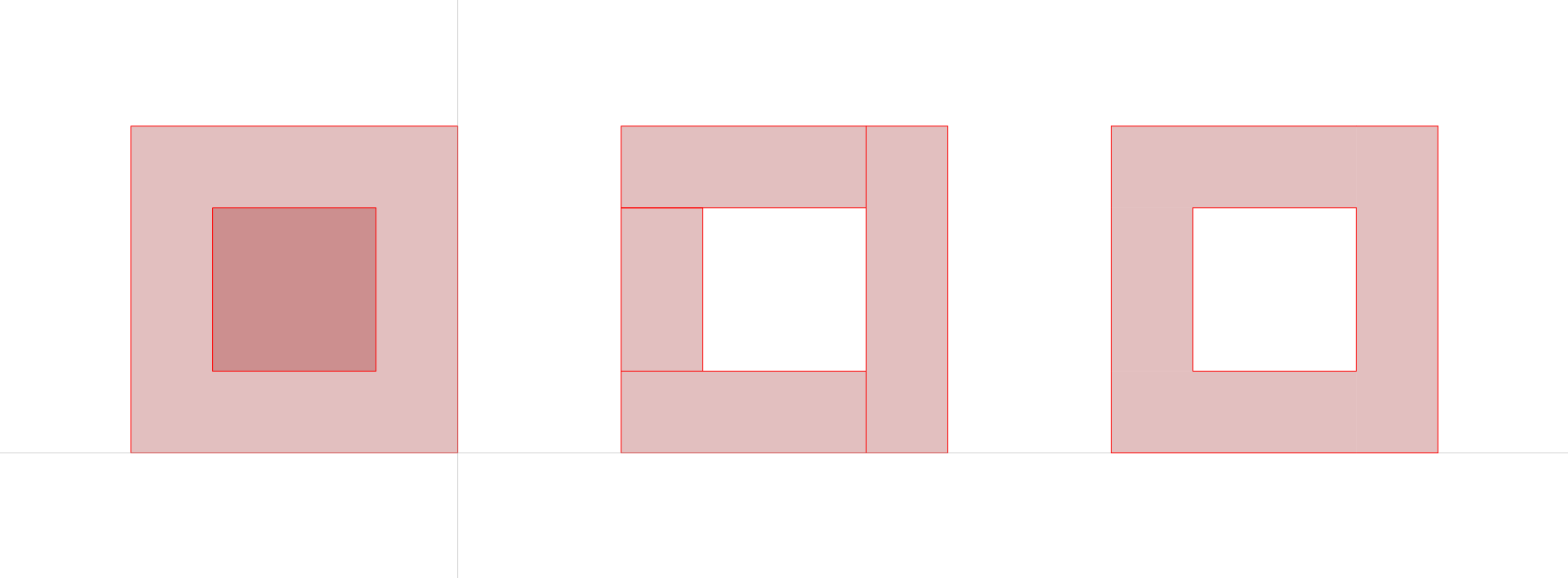}
\caption{The subtraction algorithm is illustrated on two rectangles ({\em left}).
The difference consists of four disjoint rectangles ({\em middle}), whose
outline ({\em right}) has much fewer arcs than the patches.}
\label{fig:sub}
\end{figure*}

\section{Spherical Geometry} \label{sec:geom}

With this elegant formalism in hand, the practical challenge is to derive
irreducible representations of the results of these operations,
discarding empty regions and redundant constraints. In general,
this can be a compute-intensive task but, most of the time it is very fast
and done as a pre-processing step that is well-worth the effort.
The description not only becomes more compact but the simpler form speeds
up subsequent analyses.
This region {\em simplification} is the topic of this section, where
one has to move beyond the basic 3-D concepts of the previous paragraphs
and solve the spherical geometry of the region.

\subsection{Patches and Simplified Convexes}

A region is the union of convexes, so one has to start by simplifying all
of its convexes individually.
For many of the subsequent tasks the frequency of intersections will depend on the
radii of the caps, thus it is a good practice to sort the collection by increasing
size. After making sure that the smallest cap is indeed finite (if infinitely small,
the convex is empty, and is eliminated), we examine the pairwise relations of the
halfspaces. Based on the topology of the halfspaces, we can discard the ones that
are the same as another, and those that fully contain other halfspaces. If we find
halfspaces that are each other's inverse or simply disjoint, the convex is empty.

Having done the trivial pruning of halfspaces, there might still be redundant
constraints but one can only reject them by explicitely solving for the vertices
and arcs of the convex.
%
%
The circumference of a halfspace, a circle, would generally intersect other circles.
We compute the roots (0, 1 or 2) for all possible pairs of circles, keeping track
degenerate roots where multiple ($>2$) halfspaces intersect.
We now collect the roots on each circle and form arcs around the circle between the
roots. Some of the arcs are invisible, i.e., outside the convex, these we can ignore,
and can also prune degenerate ones, if any.
At this point, one can start to form a chain (linked list) of the arcs by taking one
randomly and looking to match its end point with the start point of a next arc.
Then we repeat adding arcs until the starting point of the first arc is reached. This
singly connected part enclosed by the arcs is called a {\em patch}.

In general, a convex could have more than one patches, in which case there could be
leftover arcs. With these we can repeat the previous procedure to form the remaining
of the patches.
In some sense, the patches are the most basic and compact elements of the convexes.
We derive a minimal enclosing circle for every patch and store them with the convex.
The bounding circles facilitate quick rejections in containment tests for convexes
of many halfspaces, and enable faster collision or overlap detections between shapes.
In the process we keep track of the halfspaces whose arcs are present in any of the
patches, these are the halfspaces that we have to keep in the simplified representation.
On top of these, we have to add those halfspaces that are needed to reject the roots
outside the convex.
This may happen in situations such as the diamond shape that is defined as the
intersection of four holes. Since the surface of the sphere is a closed manifold,
these four negative halfspaces could define two patches so another halfspace is
needed to select the patch we want, see Figure~\ref{fig:diamond}.
By the end of the algorithm, the convex is left with the minimum number of halfspaces
that still describes the same shape.
%

\subsection{Region Simplification}

Naturally, the simplification of a region starts by processing its convexes
one by one. If we know that the convexes are disjoint or not interested in
the analytic area calculation, we are done.
The following steps can not only provide a region description with disjoint
convexes and precise areas but enable a potential performance boost from a simpler
representation.

Building on the bounding circles, first an approximate {\em collision graph}
is calculated, where the links note which convex (a node in the graph)
might overlap with others.
It is possible that some convexes simply contain others, in which case the
smaller ones are redundant, and thus quickly eliminated, simplifying the
collision graph.

Partial overlaps are more difficult to deal with, and here the region algebra
proves (see Section~\ref{sec:alg}) to be invaluable.
To guarantee disjoint convexes, one has to look for potential collisions
between two convexes and derive new ones that do not intersect. In practice,
this is simply the convex subtraction, where one keeps the larger convex
and substitutes the smaller one with the difference of the two.
The collision graph is updated to include the new convexes that are now 
disjoint from the larger one. Since the new convexes present a smaller area 
than their progenitor, they can only collide with those that were linked to 
the eliminated one.
We repeat this procedure until the graph has no links left. We proceed by 
working with the larger convexes to guarantee that those are not broken up 
into many small pieces. It is common for regions to have more convexes than 
the original description after this step, but they will be now disjoint.
During this simplification process the number of collisions often grows as 
for every eliminated convex we introduce one or more, whose collision links 
are inherited and get queued for analysis.

It is worth noting that one can also eliminate convexes by {\em stitching} 
them. For example, two adjacent rectangles next to one another that share an 
arc and have neighboring halfspaces that are identical, can be substituted by 
a single convex.
Stitching is not only a cosmetic improvement but in certain situations can 
make a huge impact on the performance. One example is when working with pixels 
that can be merged. Heuristic simplifications can make use of this feature,
as we will see in Section~\ref{sec:igloo}.

\subsection{The Outline of a Region}  \label{sec:outline}

The dual halfspace-convex-region representation has many advantages as discussed 
before, but in certain situations a third representation, the {\it outline}, 
provides further new opportunities.
One obvious application is visualization, when we would like to render regions 
projected on the sky. In Section~\ref{sec:htm} we will also look at how the 
outline can be used in advanced algorithms to accelerate searches for points in 
a region.
We started by stating that the surface and shape descriptions are theoretically 
equivalent but in practice going from one to the other is not always
straight forward. Next we discuss the outline and its derivation.
Going the other way, from outline to surface representation is
less convenient and not really necessary, when working primarily
with halfspaces and convexes.

The patches of a region are essentially the outlines of the convexes,
but not necessarily the outline of the region.
For example, the outline of a region with a single convex that has only
one patch consists of the patch's arcs. This is also true if the convex
has multiple patches as patches of the same convex can never touch
each other at more than a point. As soon as multiple convexes
are present, adjacent ones will share arcs, at least partially.
We can define a {\it segment} as part of a directed arc that has a start
and an end point, and there are no additional vertex point along the arc 
between these two. The algorithm to create an outline would just have to 
remove those segments from all patches that ``annihilate'' each other, i.e. 
there are two directed segments which are identical in geometry but opposite 
in direction.
The first step is to break up each arc into unique segments, which is 
performed by grouping the arcs by common circles, then ordering the 
circle, and breaking all arcs up into distinct segments.
A possible approach to identify the ``canceling'' segments is to look along
the common circles in the collection of all patches and for every circle
identify the places where two segments are opposite to one another, and 
hence cancel out. In this case both are removed.
To preserve the relation to the original patches, which have precomputed
bounding circles, we keep the structure of the outline similar to that of the
region, but the arcs of the patches are replaced by smaller segments. Such
representation is most advantagous but lacks the connectivity of the
arcs, which requires the visualizer to draw the arcs independently lifting
the pen. A continous outline is chained into a loop of the correct order
by checking the start- and endpoints.

\subsection{Heuristic Simplification with Igloo} \label{sec:igloo}

When none of the above simplification methods work well, we can use hybrid
techniques where we combine the advantages of pixelization with the exact
geometry.
Let us imagine an large area survey that takes tens of thousands of
pointed observations with a circular field-of-view, so that the circles
fully overlap. While the edges of the survey are rippled and need many circles
to represent it accurately, the inside of the region is contigous and simple.
The aforementioned simplification rules including the stitching will
not be able to achieve much improvement, although there should be a simpler
form. The previous algorithms would eventually also choke on a region with
$\sim$50,000 convexes where every one of the convexes overlaps with 12 other.
This is similar to a good approximation for the sky coverage of the Faint
Images of the Radio Sky at Twenty-centimeters \citep[FIRST;][]{first} survey.

If one could break up the region into small pieces that, unlike small circles,
can be stitched together neatly, then one could potentially define a small
number of large convexes in the middle of the region, and eliminate the union of
tens of thousands of caps.
The algorithm is an elegant divide-and-conquer recursion.
We need a pixelization that is a hierarchical subdivision of the surface of
the sphere, where the pixels are not only nicely adjacent but also share many
halfspace constraints that allow for extensive merging. Such pixelization is
achieved by the Igloo scheme \citep{igloo}.
We start with a region and build a tree, e.g, using the (3:0:3) pixelization,
where only those nodes are kept that collide with the region.
Every Igloo pixel is a simple convex, aswe have defined this term, namely a
generalized spherical triangle or a rectangle, hence the usual region
operations can be directly applied.
The depth-first recursion is very efficient when one records for every node of
the tree the convexes of the region that it overlaps with, so that on the deeper
levels of the tree with more and more nodes, less and less convexes are to be
tested per each node.
We typically specify the stopping criteria for building the tree by pixel size
(maximum number of levels) or by the maximum number of leaf nodes.
Once the tree is ready, the geometry of every leaf is examined. We subtract the
colliding convexes of the region from each leaf one by one and calculate its
area (see Section~\ref{sec:area}).

If the pixel is completely covered, we take the convex of the pixel, otherwise
derive its exact shape.
The merging of the pixels into larger and larger convexes is done along the tree
using the full pixels only, and once the fragmentation of the region stops the
pixels to grow, one can stitch the neighboring pixels, if they are on the same
level.

This procedure yields the exact region with alternative internal convexes. Taking
it one step further, we can also create an even simpler region by keeping the full
pixels for nodes where the overlap area is large even if not completely full.
For these nodes, we define another region called the {\em mask} that defines the
overshoot area of the approximation.
Typical footprints already have such masks to censor areas covered by bright stars,
satellite trails, etc., and adding some more does not make the subsequent analyses
more difficult, instead, it can significantly simplify the description of the
region.

\begin{figure*}
\epsscale{1.15}
\plottwo{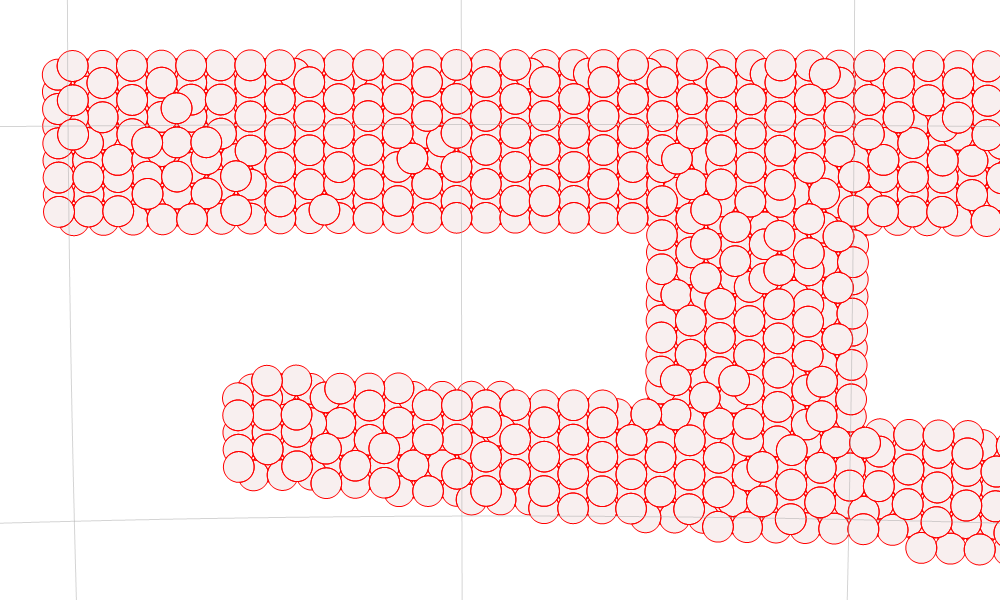}{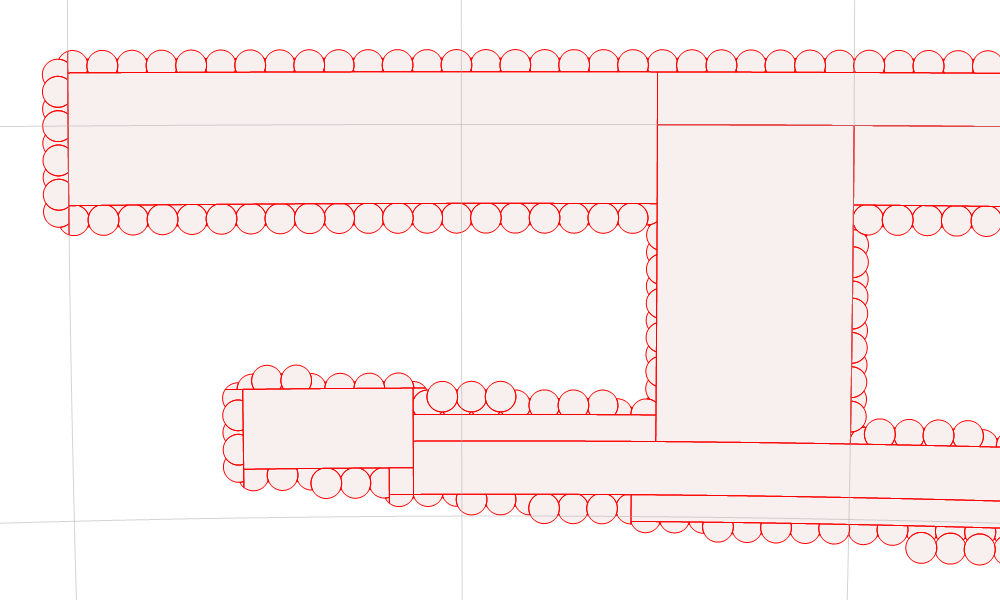}
\caption{Part of the FIRST footprint illustrates the power of the heuristic
simplification based on a hierarchy of Igloo pixels.
The brute-force method ({\em left}) runs much slower and produces many more disjoint
convexes than the novel technique ({\em right}) that fills the inside of the region
with stitchable convexes and merges them well.}
\label{fig:first}
\end{figure*}

\subsection{Area Calculation} \label{sec:area}

A patch is just a generalized spherical polygon bounded by small (and/or great)
circles, whose arcs are ordered by design. The trick is the break a patch up into
more regular pieces whose areas we can calculate.
Let us pick a point arbitrarily on the sphere, e.g., at the center of mass
of the vertices of a patch, and break up the polygon into spherical triangles
such that one of the vertices of every triangle is this selected point
and it contains an arc of the patch.
Now every triangle has two great circle arcs and one of the original arcs.
We follow the approach of \citet{goodchild92}, by subdividing this shape into
a triangle of great circle arcs and the leftover shape, known as the {\em semilune}.
The area of the former is given by the Girard formula that has several variants.
Here we list the one that is the most robust against degeneracy and hence
should be preferred for numerical calculations over equations that appear to be
simpler algebraically:
\begin{equation}
A_G = 4 \tan^{-1} \sqrt{z}
\end{equation}
with
\begin{equation}
z =  \tan\left(\frac{s}{2}\right)
\tan\left(\frac{s\!-\!a}{2}\right) \tan\left(\frac{s\!-\!b}{2}\right) \tan\left(\frac{s\!-\!c}{2}\right)
\end{equation}
where $s$ is half the circumference of the triangle with the sides $a,b,c$ in radians.

The area of a semilune is also calculated analytically but it is more convoluted.
For completeness, here we print the formulas without further explanation\citet{goodchild92}.
\begin{equation}
A_S = a - b\cos\theta
\end{equation}
where $\theta$ is the half angle of the small circle, i.e., the radius of the cap,
and
\begin{eqnarray}
a & = & 2 \arcsin\left(\tan\left(\arcsin r\right) \big/ \tan\theta\right) \\
b & = & 2 \arcsin(r\,/\sin\theta)
\end{eqnarray}
with $r$ being half the Eucledian distance between the end points of the arc.
The situation is slightly complicated by the fact that certain patches are actually
holes (for \mbox{$\cos\theta<0$}), whose area is propagated with a negative sign
to obtain the correct total sum for the region.

\iffalse

\begin{figure*}
\epsscale{1.2}
\plotone{fourlevels.png}
\caption{The recursive subdivision of the octahedron, illustrated above,
is at the heart of the Hierarchical Triangular Mesh, also known as the HTM.}
\label{fig:subdivision}
\end{figure*}

\else

\begin{figure*}

\includegraphics[width=35mm, trim=110 110 110 110, clip]{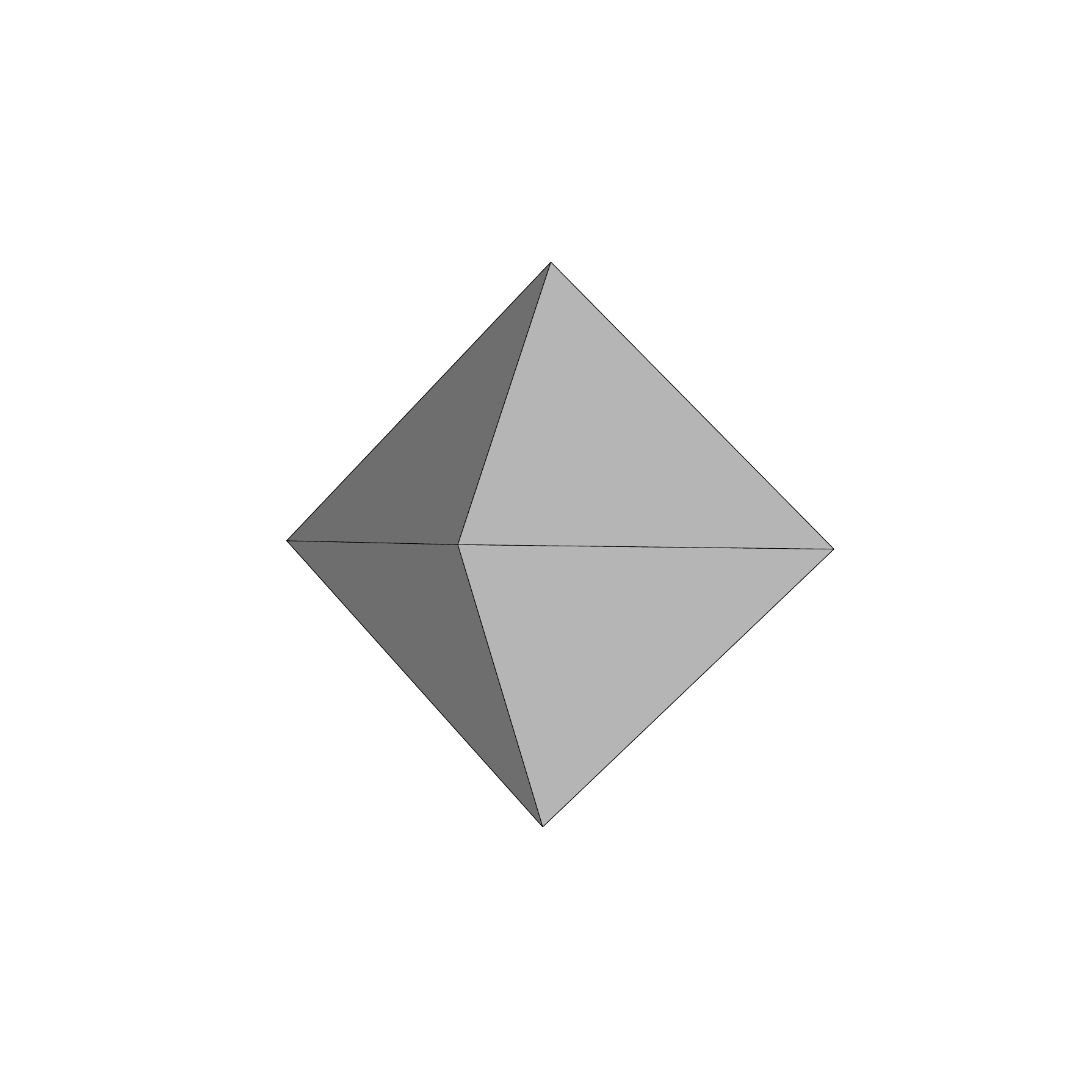} \hfill
\includegraphics[width=35mm, trim=110 110 110 110, clip]{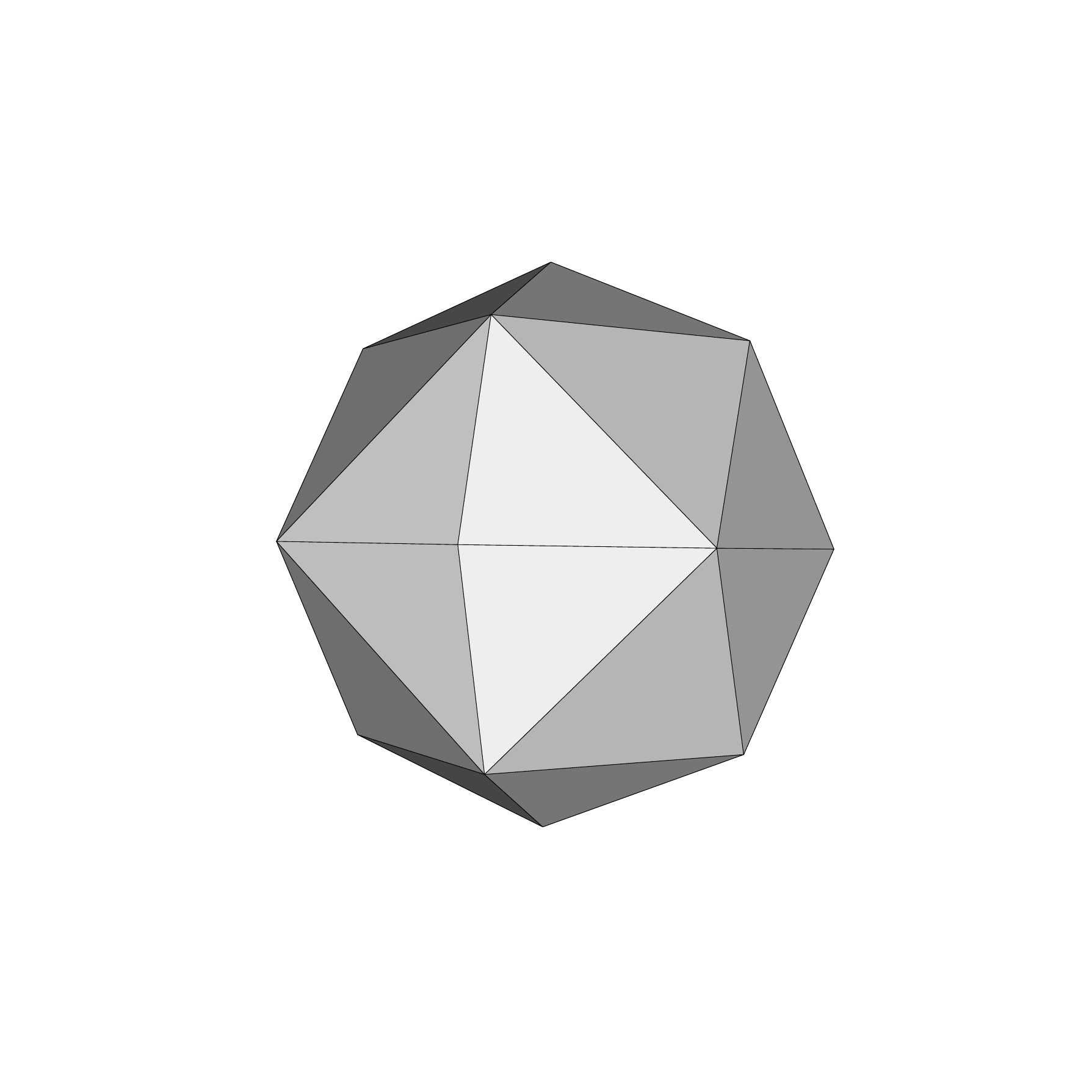} \hfill
\includegraphics[width=35mm, trim=110 110 110 110, clip]{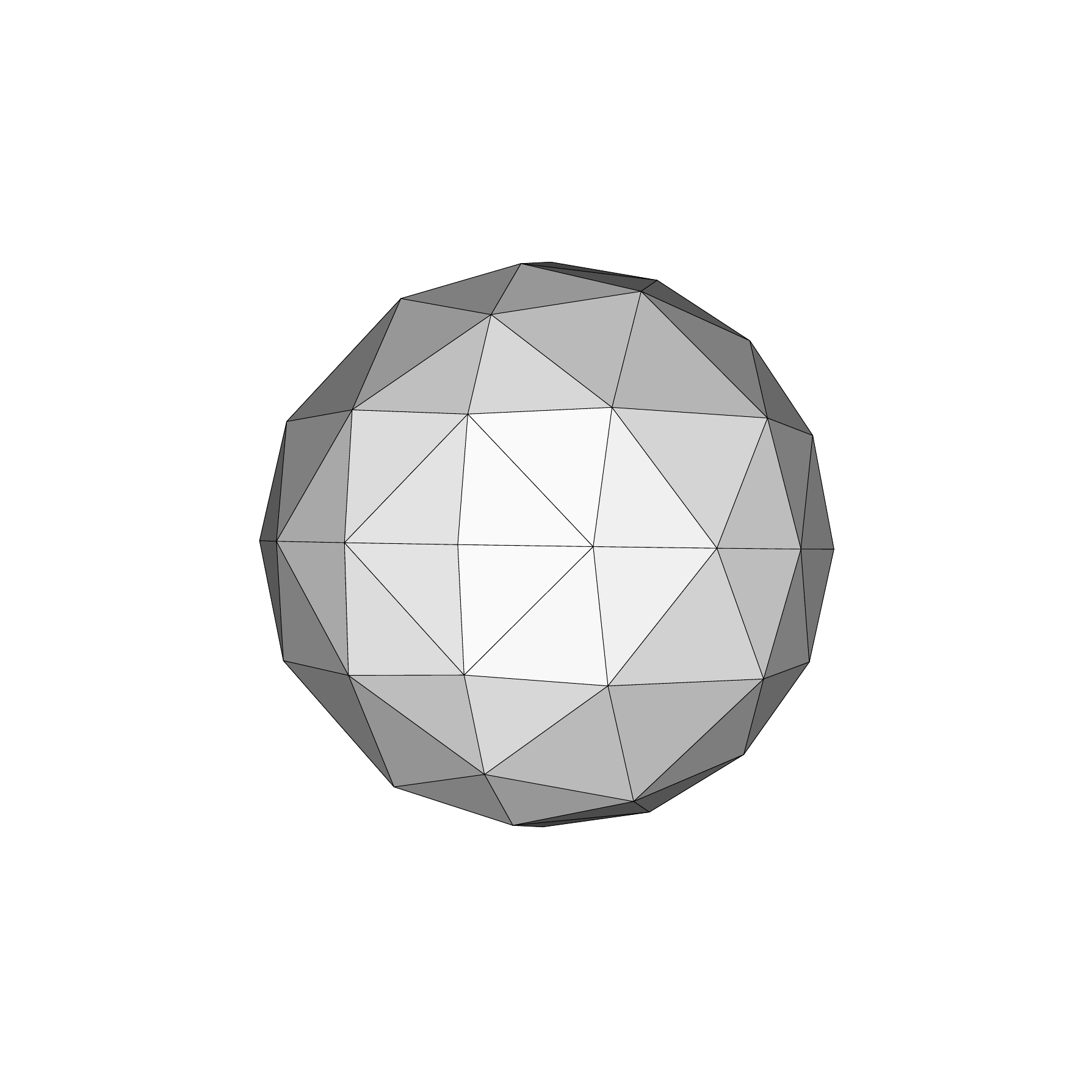} \hfill
\includegraphics[width=35mm, trim=110 110 110 110, clip]{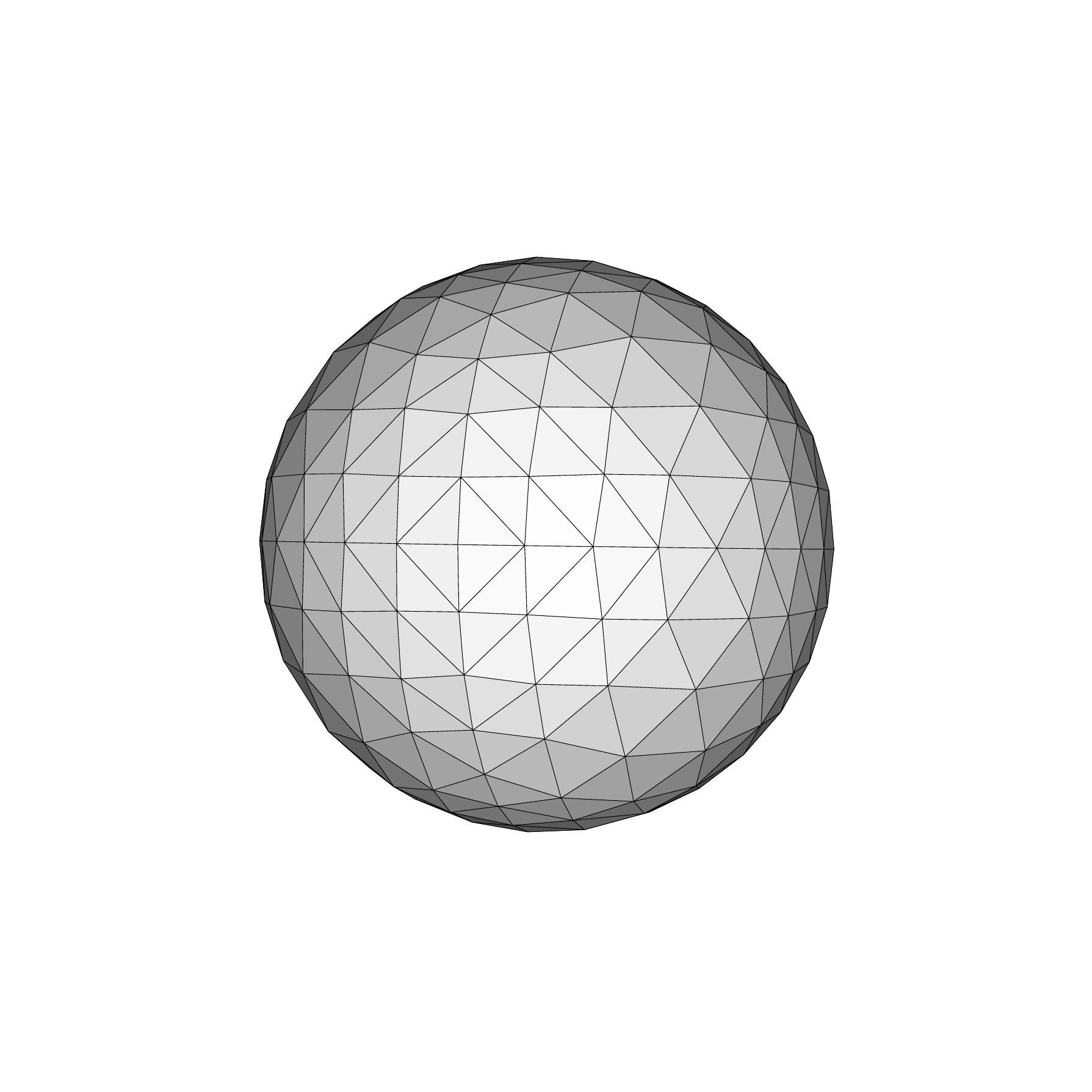} \hfill
\includegraphics[width=35mm, trim=110 110 110 110, clip]{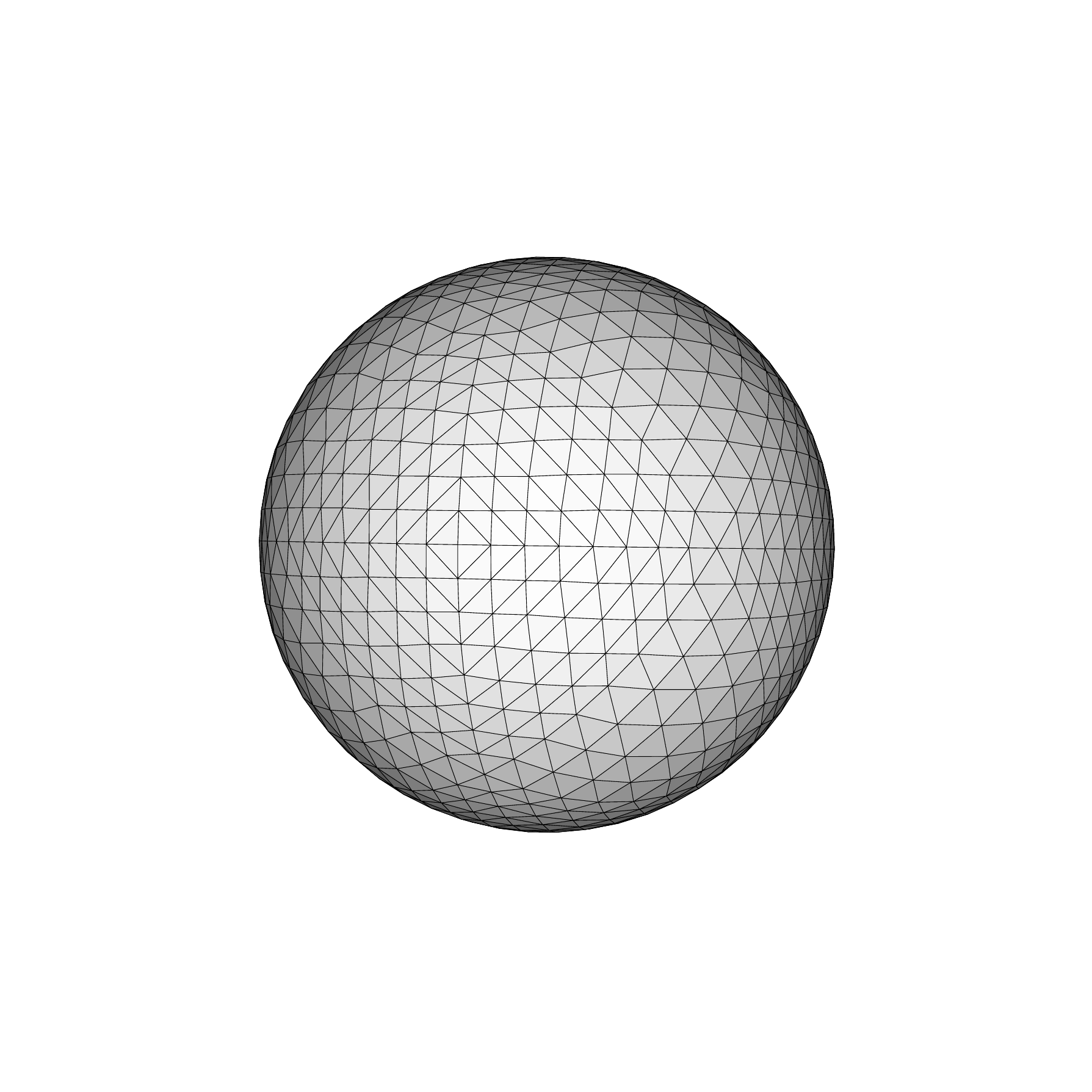}

\caption{The recursive subdivision of the octahedron, illustrated above,
is at the heart of the Hierarchical Triangular Mesh, also known as the HTM.}
\label{fig:subdivision}
\end{figure*}

\fi

\section{The Hierarchical Triangular Mesh} \label{sec:htm}

Another pixelization of the sphere in combination with the
region representation enables fast spatial searches for
catalog entries of stars, galaxies and other astronomical sources
that are within a spherical region on the sky.
The idea is to apply a coarse filter to the entire dataset
and to reject most of the sources that are outside the search region,
and perform the computationally more expensive geometry test on a much
smaller subset of candidates that already passed the filter.

Our choice for such a filter is based the Hierarchical Triangular Mesh,
also known as the HTM \citep{kunszt00}. Next we discuss the properties and
features of the HTM, then introduce the algorithms for creating efficient
coarse filters for regions that map very well onto the indexing facilities
of currently available relational database engines.

\begin{figure}
\epsscale{0.8}
\plotone{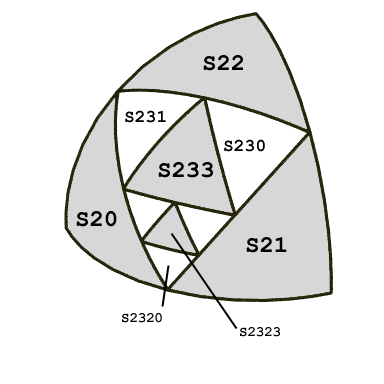}
\caption{The naming of the hierarchical trixels provides unique identifiers.
The ensemble of all trixels at a given level can be thought of as a
space-filling curve on the surface of the sphere.}
\label{fig:naming}
\end{figure}

\begin{figure*}
\includegraphics[width=42mm]{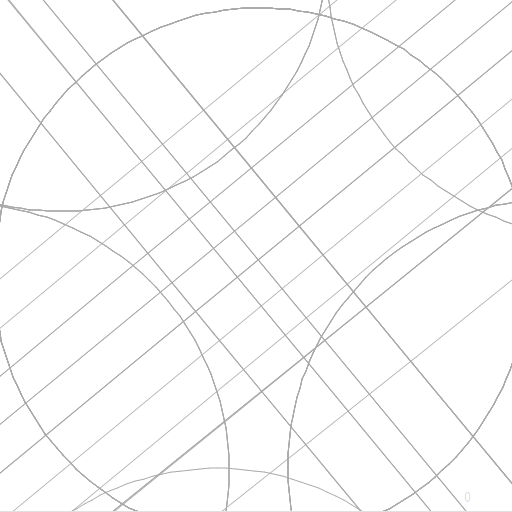}\hfill
\includegraphics[width=42mm]{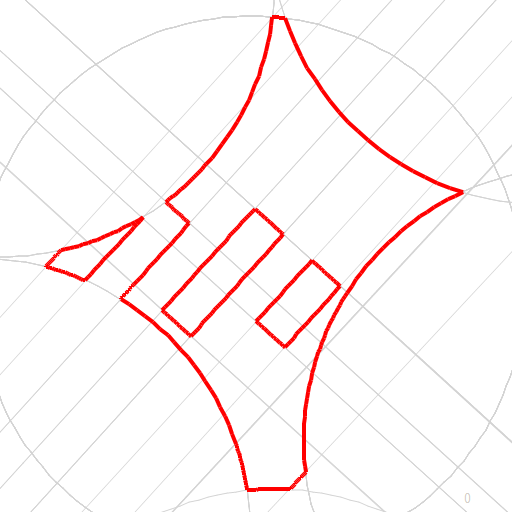}\hfill
\includegraphics[width=42mm]{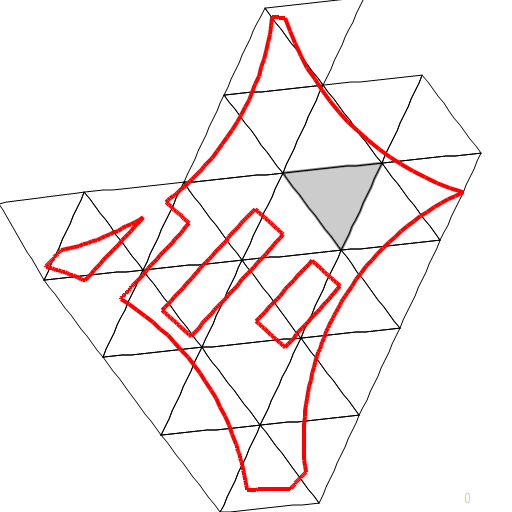}
\includegraphics[width=42mm]{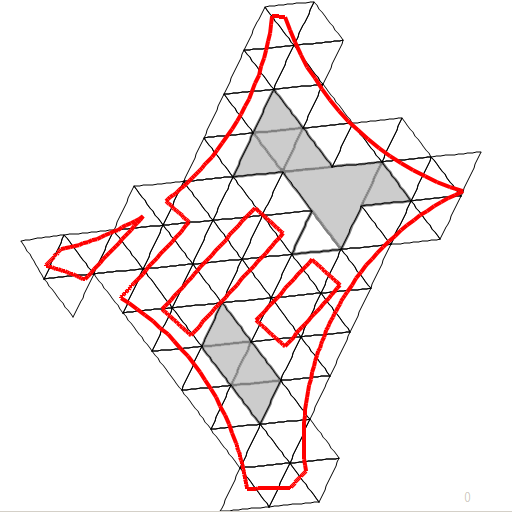}
\caption{A sample region from the SDSS geometry illustrates the various
spherical concepts: ({\em from left to right}) (1) Circles of all halfspace contraints
of the region. (2) The outline as used in the HTM algorithm. (3) The inner ({\em filled gray})
and outer covers ({\em open triangles}) of the region using larger and (4) smaller trixels.}
\label{fig:R45}
\end{figure*}

\subsection{Address of a Point}

We can paint the sphere with triangular pixels that we call
{\em trixels} defined by the HTM.
The top nodes are the 8 faces defined by an octahedron projected onto
the sphere. The children of each node are obtained by subdividing the
existing triangular nodes into 4 new triangles. The sub-triangles have
the current corners and the current arc's midpoints as their corners.
Finer detail is created as new levels are added by repeating the process
for each triangle. In the limit, the recursive subdivision approaches
the ideal sphere as in Figure~\ref{fig:subdivision}.

All the triangles at any level have a unique identifier or trixel ID.
It is an integer number, that encodes the position of the trixel in
the hierarchy and is composed through the following recursive algorithm.
The level 0 trixels are the faces of the octahedron.
We name them with {\tt N0}, {\tt N1}, {\tt N2}, {\tt N3} and
{\tt S0}, {\tt S1}, {\tt S2}, {\tt S3} where {\tt N} and {\tt S}
refer to the northern and southern hemispheres, respectively.
In the recursion, each triangle has four offsprings
that are named by appending the index 0, 1, 2 or 3 to the parent's name.
Figure~\ref{fig:naming} illustrates the naming convention in the
subdivision on a few examples.

To keep things consistent, we introduce a straightforward mapping between
the name and the ID of a trixel.
The ID is a 64-bit integer and is more compact than the human-readable
name that can be up to 25 bytes long.
First we assign the bits {\tt 11} to {\tt N} and {\tt 10} to {\tt S} and
convert the trixel's number betwen 0 and 3 to binary ({\tt 00} and {\tt 11})
and append it to the previous bits. We repeat until the desired level is reached.
For example, the trixel named {\tt S2320} will convert to binary {\tt 1010111000}
or decimal 696. The longer the name of a trixel, the deeper it is in the hierarchy.
For practical purposes, we stop at level 20, approximately corresponding
to a positional accuracy of about 0.3". This special level-20 trixel ID is called
the {\tt HtmID} in our system. A 64-bit integer can hold a trixel number to level
30, although the standard IEEE-float double precision breaks down after level 25 where
the positional accuracy would be about one hundredth of an arcsecond.
A property of the trixel ID numbers is that the descendants of a trixel have IDs
that form a consecutive list of numbers. For example, the 4 children of {\tt S2320},
namely the set of \{{\tt S23200}, {\tt S23201}, {\tt S23202}, {\tt S23203}\},
form the range of numbers 2784--2787.
The level 20 offsprings of the same trixel form the HtmID range
11957188952064--11974368821247.
This is important, because we can represent any level trixel with either a single
number, or with a pair of low-high HtmID values. The consequence of using the latter
representation is that regions with variable size trixels can be expressed uniformly,
and because the numbering provides partial coherence of the HtmID numbers.

Since trixels partition the sphere, any location is inside exaclty one trixel, so
a level 20 trixel ID, or the HtmID is a fairly accurate approximation of the
position. This property is exploited in our spatial search algorithm detailed
in the subsequent paragraphs.

\subsection{Approximate Region Covers}

Simply put, a {\em cover} is a set of trixels that fully covers a region.
It is an approximation of the shape by the union of a set of spherical triangles.
Given a region, the algorithm starts with the eight level-0 trixels that
make up the entire globe. The {\tt S2} in Figure~\ref{fig:naming} is one of these
eight trixels that are initially marked as unprocessed.
In the recursion, all unprocessed trixels are analyzed and
get marked with one of three possible tags.
The {\em inner} trixels are fully inside the region,
the {\em reject} ones are completely outside and
{\em partial} trixels are on the outline.
Dealing with inner and reject trixels is easy. Inner trixels are
saved for output, and the rejects are discarded from further consideration.
A partial trixel is subdivided into four smaller trixels which are
placed at the back of the unprocessed list.
Eventually all trixels are tagged to the desired level of detail.

Often it is very useful to have an approximation of the inside
of the region that does not touch the outline.
Sources in the inner cover are guaranteed to be inside the region, hence 
do not require extra geometry tests.
This {\em inner cover} is the union of all the inner trixels,
while the aforementioned {\em outer cover} is the union of
the inner and partial trixels.
In fact, it is possible to obtain both sets at the same time
without much extra processing.
As to where to stop, there is no universal optimum, and the answer
will depend on the actual dataset and region, as well
as the implementation of the search engine and even
its hardware configuration. Fortunately, sensible heuristics
exist and the performance of the searches are considerably better
than the naive implementation for any reasonable cover shapes.

\subsection{Searching with HTM}

Let us now examine how spatial searching is performed
using an HTM cover.
Assume that every source of our dataset has a pre-computed HtmID 
that constrains its location on the celestial sphere to within a 
particular level-20 trixel.
The outside cover of the region can be represented as a
set of HtmID intervals. If the sources are ordered by their
HtmID, fetching sources in these intervals is extremely fast.
If the dataset resides on disk, which is the case for any massive
live astronomical catalog, getting sources in an interval consists of
reading sequentially from the hard drive. At the end of every interval,
the disk head is raised and re-positioned to the beginning of the next 
interval. This seek-time is the source of the penalty one would pay
for a very accurate cover that is represented by many short intervals.
Often a couple of dozen HTM ranges provide accurate representations 
that select a small enough candidate list with which modern CPUs can 
keep up with processing the exact geometry calculations.
In general, any custom stopping criterion can be utilized, e.g.,
based on elapsed time or resolution size.
One particularly interesting possibility is to monitor the area of
the inner and outer covers and stop at a desired limit on their ratio.
In fact, the area ratio can be approximated by the ratio of the number trixels
in the two covers. The areas of trixels on the same level vary less than about 40\%
but the collections of (random) trixels would average out this variance
to a more precise estimation of the area ratio.

\section{Software Packages} \label{sec:lib}

Our design of the software implementation was driven by
several requirements. It needed to be architecture
and operating-system independent that also integrates
well with the relational database technology, which is
at the core of most modern astronomy archives today.
We chose to build our solution in the .NET Framework%
\footnote{Click to visit {\url{http://microsoft.com/net}}}
programming model that
satisfies our development, maintainance, portability and performance
requirements.
Such managed code runs in a virtual machine called
the Common Language Runtime, or CLR for short.
In addition to Microsoft's CLR implementation of the Common Language
Infrastructure (CLI), there is also an
open-source cross-platform runtime by
the Mono Project%
\footnote{Click to visit \url{http://www.mono-project.net}}.
The .NET Framework is not only OS independent but
also allows for development in several programming languages
and the integration of projects in a mixture of languages.
Using the C\# programming language, we built a set of class libraries
that depend on each other, so that
applications in any one of the supported languages 
can choose to include the appropriate modules selectively.

\subsection{The Spherical Library}

The basic module or assembly contains the routines to deal with
the spherical geometry. Generic container classes are used to store
the collection of halfspaces in a convex and the collection of convexes
that make up a region. Thus managing a shape is as simple as working
with lists.
Here we show the C\# listings of a trivial example with a convex of 
two halfspaces to illustrate the simplicity of the coding.
First, we define the centers of the caps using J2000 (R.A., Dec.) coordinates,
\begin{center}\begin{minipage}{8cm}{\footnotesize\begin{verbatim}
Cartesian p1 = new Cartesian(180, 0);
Cartesian p2 = new Cartesian(181, 0);
\end{verbatim}}\end{minipage}\end{center}
then set the radius of the caps and create the halfspaces
using the centers
\begin{center}\begin{minipage}{8cm}{\footnotesize\begin{verbatim}
double theta = Math.PI / 180;    // 1 degree radii
Halfspace h1 = new Halfspace(p1, Math.Cos(theta));
Halfspace h2 = new Halfspace(p2, Math.Cos(theta));
\end{verbatim}}\end{minipage}\end{center}
The convex is a collection of halfspaces that are added
one by one after which we invoke the simplification of
the description that also derives the arcs and patches of the
shape along with its analytic area in square degrees.
\begin{center}\begin{minipage}{8cm}{\footnotesize\begin{verbatim}
Convex c = new Convex();
c.Add(h1);
c.Add(h2);
c.Simplify();
Console.Out.WriteLine(c.Area);
\end{verbatim}}\end{minipage}\end{center}
Similary a region is created by adding convexes to its collection.
\begin{center}\begin{minipage}{8cm}{\footnotesize\begin{verbatim}
Region r = new Region();
r.Add(c1);
r.Add(c2);
r.Simplify();
Console.Out.WriteLine(r.Area);
\end{verbatim}}\end{minipage}\end{center}
Naturally, the number of convexes in a region and halfspaces in a 
convex are not limited to two (and can also be one), they are only 
bound by system memory and computational power for processing.

Boolean operations on the shapes are implemented to perform the 
computations in place whereever possible. For example, the union of 
regions or the intersection of convexes can be done within the instance 
on which the method is called, but the difference of two convexes or 
regions is implemented to return the resulting region.
On top of the straightforward translations of the algorithms in Section~{sec:alg},
most operations have a {\em smart} version for simplified shapes. One of the
advantages of using these methods is the speedup for complicated shapes, where
the precomputed bounding circles reduce the computational costs.
The other benefit is potentially even greater.
E.g, the union of simplified regions can be done based on the assumption that
the convexes of the regions are already disjoint, and hence, check for collisions
among the mixed pairs only. In code, the snipets
\begin{center}\begin{minipage}{8cm}{\footnotesize\begin{verbatim}
r.Union(r2);      // regions may not be simplified
r.Simplify();     // simplification from scratch
Console.Out.WriteLine(r.Area);
\end{verbatim}}\end{minipage}\end{center}
and
\begin{center}\begin{minipage}{8cm}{\footnotesize\begin{verbatim}
r.SmartUnion(r2); // both simplified
Console.Out.WriteLine(r.Area);
\end{verbatim}}\end{minipage}\end{center}
are conceptually identical but the latter can take significanly less
time as it can assume the arguments to be in the canonical form.

The application programming interface (API) supports many more advanced ways to
create and manipulate regions of arbitrary complexity, the working code snipets
above are here to serve as guide.
%

\subsection{Numerical Imprecisions}

Numerical stability of the implementation of the spherical algorithms is
crucial and has not been easily achieved. Computations with floating-point numbers
make errors that can accumulate in a series of operations.
These uncertainties can result in erroneous determination
of topological relations of points and planes, convexes and regions.
One possible solution is not to use floating-point arithmetics. Software
packages exist that work with rational numbers, and represent them
as a pair of integers: the numerator and the denominator. In this setting,
the values are always precise but there is an efficiency penalty.
In spherical geometry, there is an even more severe problem with this approach.
The set of rational numbers is not closed for operations that one has to
routinely perform, e.g., the square root of a rational number can be irrational.

We use IEEE Standard 754 double-precision floating-point numbers in our
implementations that most modern CPUs can efficiently process, and we
carefully analyze the code for numerical stability and rounding problems.
It is a surprisingly hard task. For decades, linear algebra routines, e.g., those in 
LAPACK, have been strengthened and optimized by hand because no generic solution 
exists; only best practices. Classic examples include the robust solution to the 
quadratic equation, or evaluating the values of $a^2\!-\!b^2$ and $\log(1+x)$.
To illustrate the importance, here we describe real-life problems
that make the spherical geometry computations more difficult than expected,
and explain our solutions briefly. The details of error propagation
in floating-point arithmetics is beyond the scope of this writing.
Rather, we refer the interested reader to \citet{float}.

Previously we said that deciding whether a point is inside a halfspace,
the most basic containment test discussed in Section~\ref{sec:shapes},
involves calculating the dot product of the point's unit vector and
the direction of the plane and comparing the result to the halfspace's
offset, the cosine of the cap's radius. The numerical imprecision on
the result translates to larger errors in the radius because the
cosine function is quadratic at values near zero, hence
using the sine can be more advantageous.
Even then, when the point is very close to the plane, it is essentially
impossible to decide which side it is on or whether it is contained
in the plane. We circumvent the problem by not trying to answer
in or out but allow for an {\em undetermined} value within a margin
derived from the limitations of the representation of floating-point 
numbers. This value is deep in the core of routines establishing the 
spatial topological relations of the shapes.

Another, more subtle, algorithmic problem is the solution of the
intersection of two planes, and the derivation of the intersecting
points of two circles on the unit-sphere.
%
%
As discussed in detail by \citet{priamos}, an elegant robust solution
(out of many possible derivations) is based on optimizing for the errors
made in the computation of a point that the line crosses.
The idea is to pick one of the \mbox{x-y}, \mbox{y-z} and  \mbox{x-z} 
planes that is most perpendicular to the direction of the line, and
solve for the point in that plane. This way one of the coordinates
is readily fixed to be 0, and we only have to solve a 2-D linear
equation for the other coordinates.

Halfspaces and vertices of spherical polygons become degenerate
very frequently as a result of routine operations while building up
the geometric description of observations.
Without controlled accuracy these degeneracies
could not be spotted and caught to derive correct representations
or the regions.

\subsection{Modules for HTM}

Classes and routines of the HTM implementation are divided into
two basic categories.
The creation of the hierarchy of triangles is a module of its own.
The tree is usually not created for its extremely large size
but computed on the fly. Once the algorithm is fixed, the hierarchical
trixels exists without the actual tree in memory or disk.
The methods provide the recursion and the basic geometrical description
of the pixels. For example, a single call yields the HtmID,
\begin{center}\begin{minipage}{8cm}{\footnotesize\begin{verbatim}
Cartesian p = new Cartesian(180,0); // Point on the sky
Int64 htmID = Trixel.CartesianToHid20(c); // and its ID
\end{verbatim}}\end{minipage}\end{center}
and another the trixel's geometry,
\begin{center}\begin{minipage}{8cm}{\footnotesize\begin{verbatim}
Cartesian v1, v2, v3;   // Vertices of the trixel
Trixel.ToTriangle(htmID, out v1, out v2, out v3);
\end{verbatim}}\end{minipage}\end{center}

On top of the core module is a set of classes that deal with the regions
and their topological relations to the trixels.
Our efficient implementation of the HTM makes use of
the internal structure of the region, and the derived properties of its
convexes and patches as well as the outline.
In the API, all the complexity is hidden behind simple
\begin{center}\begin{minipage}{8cm}{\footnotesize\begin{verbatim}
List<Int64> trixels = Cover.HidList(region);
Console.Out.WriteLine(trixels.Count);
\end{verbatim}}\end{minipage}\end{center}
Alternatively one can explicitly instantiate a cover
object, and investigate the properties during
processing
\begin{center}\begin{minipage}{8cm}{\footnotesize\begin{verbatim}
Cover k = new Cover(region);
k.Run();  // Default processing and stopping
          // Call k.Step() for more control
List<Int64> inner = k.GetTrixels(Markup.Inner);
List<Int64> outer = k.GetTrixels(Markup.Outer);
List<Int64> partl = k.GetTrixels(Markup.Partial);
\end{verbatim}}\end{minipage}\end{center}
Similarly the intervals are retrieved by a single method call
and print in the following example.
\begin{center}\begin{minipage}{8cm}{\footnotesize\begin{verbatim}
List<Int64Pair> ranges = Cover.HidRange(region);
foreach (Int64Pair p in ranges)
    Console.WriteLine(p)
\end{verbatim}}\end{minipage}\end{center}

As shown in the examples above the HTM and Spherical Library
classes and routines work together seemlessly by design.
They leverage all the information available to perform
the operations as fast as possible, while keeping the usage
patterns simple. Under these high-level routines, powerful
methods provide easy customization of the code for
other type of problems and applications.
One such example is the NVO's Footprint Service that
uses high-resolution HTM ranges to look for overlapping
footprint, or regions that contain a given point.

\subsection{Region in a String}

Our basic internal string representation of the regions follows the simple
structure of the collections. We enumerate the convexes and all their
halfspaces
\begin{center}\begin{minipage}{8cm}{\footnotesize\begin{verbatim}
REGION
    CONVEX CARTESIAN x1 y1 z1 c1
           ...
           CARTESIAN xN yN zN cN
    CONVEX ...
\end{verbatim}}\end{minipage}\end{center}
with arbitrary whitespace and linefeed characters.
The cartesian coordinates are the usual transformations of the J2000
$(\alpha,\delta)$ by equations
\begin{eqnarray}
x & = & \cos\delta\,\cos\alpha \\
y & = & \cos\delta\,\sin\alpha \\
z & = & \sin\delta
\end{eqnarray}
On top of this simple description, the interfaces also support more
advanced concepts via region parser based on our grammar
in the Backus--Naur Form (BNF).
The use of these constructs are
often more straighforward than spelling out the $(x,y,z)$ coordinates
of the normal vectors. Here we show the case for a couple of
simple convexes often used by astronomers. The region describes
the union of
a circle with a radius of $60'$ around the center at $(\alpha,\delta)=(180^{\circ},0^{\circ})$
in the J2000 coordinate system
and a great-circle polygon specified by its ordered vertecies given by the angles.
\begin{center}\begin{minipage}{8cm}{\footnotesize\begin{verbatim}
REGION
    POLY J2000 180 0 182 0 182 2 180 2
    CIRCLE J2000 180 0 60
\end{verbatim}}\end{minipage}\end{center}
The polygon can be built up by any number of vertices (greater than two)
and the region can have any combination of convex constructs.
Another useful feature is the convex hull of a point set that is
specified by the keyword {\tt CHULL}.
To specify cartesian coordinates, one can replace the {\tt J2000}
keyword with {\tt CARTERSIAN} and enumerate the components of the
unit vector instead of the anglular coordinates.

\subsection{SQL Routines} \label{sec:sql}

One of the most attractive advantages of developing in the .NET programming 
model is the elegant integration with SQL Server since the 2005 Version. The 
runtime is actually hosted inside the engine, which allows for the
customization of the database to satisfy the astronomy specific requirements. 
The custom assemblies can be loaded to be part of the database along with the 
catalog data, and User-Defined Functions (UDFs) can wrap the functionality of 
the managed code that are invoked efficiently from SQL at query time. This way 
custom programs can run inside the database right there where the data are, 
and perform analyses without moving the bits on the network or even outside 
the SQL engine.

Our harness for the spherical implementation is schema-independent by design.
This means that the same SQL routines can be present and be used in any
astronomy science archive regardless of the layout of the database or the
content of the tables.
In fact, it is currently in use in a wide variety of services, including the
Sloan Digital Sky Survey, the Galaxy Evolution Explorer and the Hubble Legacy 
Archive, being integrated with the Spitzer and Chandra servers, among others.

The regions are serialized into a compact binary format that contains both
the halfspace-convex-region representation and the patches along with their
minimal enclosing circles.
The simplest way to create a region is by using the internal specification language
that can describe any region but the operations are also supported when
starting with basic building blocks. For example, the union of a 60' radius
circle and the spherical rectangle can be coded as follows.
\begin{center}\begin{minipage}{8cm}{\footnotesize\begin{verbatim}
DECLARE @s VARCHAR(MAX), @r VARBINARY(MAX),
        @z VARCHAR(MAX), @u VARBINARY(MAX)
SELECT @s = 'REGION CIRCLE J2000 180 0 60',
       @z = 'POLY J2000  180 0  182 0  182 2  180 2',
       @r = sph.fSimplifyString(@s),
       @u = sph.fUnion(@r,sph.fSimplifyString(@z))
SELECT sph.fGetArea(@r), SELECT sph.fGetArea(@u)
GO
-- 3.14151290574491  6.35572804450646
\end{verbatim}}\end{minipage}\end{center}
In SQL, the binary blobs of arbitrary size (up to 2GB) are represented
as {\tt VARBINARY(MAX)}, and here we use variables of that type. Naturally,
one can also create tables with VARBINARY columns to save the resulting
regions within SQL Server.
We note that previous version of the SQL harness, e.g., currently deployed in
the SDSS Catalog Archive Server, do not use the separate {\tt sph}
schema but instead keep the UDFs in the default {\tt dbo} using an {\tt Sph} prefix,
e.g., in {\tt dbo.fSphGetArea(@r)}.

The HTM routines in SQL provide high-performance spatial searches in combination
with the builtin indexing mechanism.
Scalar-valued UDFs can compute the address of a point
at the default level of 20. Here we list a simple example that updates a table
called {\tt PhotoObj} to set the {\tt HtmID} column for all rows based on
their J2000 coordinates
\begin{center}\begin{minipage}{8cm}{\footnotesize\begin{verbatim}
UPDATE PhotoObj SET HtmID = dbo.fHtmEq(RA,Dec)
\end{verbatim}}\end{minipage}\end{center}
Using spherical regions to search for sources is also very
straightforward and very efficient. In the following snippet
we fetch only the rows that are in the HTM cover, hence
are probably contained in the region:
\begin{center}\begin{minipage}{8cm}{\footnotesize\begin{verbatim}
WITH Cover AS
(
    SELECT * FROM dbo.fHtmCoverRegion
          ('REGION CIRCLE J2000 180 0 10')
)
SELECT o.ObjID
FROM PhotoObj AS o INNER JOIN Cover AS c
    ON o.HtmID BETWEEN c.HtmIDStart AND c.HtmIDEnd
GO
\end{verbatim}}\end{minipage}\end{center}
The UDF returns a table of the HTM ranges of the approximation of
the specified shape that overshoots, and the join
returns instantly all the possible rows. In a real-life search,
the {\tt WHERE} clause of the query would have a separate containment
test with the region geometry. The sole purpose of the HTM cover is to
fetch the good candidates only from the disk, and it does this very fast.

\subsection{The Astronomical Data Query Language} \label{sec:adql}

As part of the current standardization efforts of the International Virtual
Observatory Alliance%
\footnote{Click to visit \url{http://www.ivoa.net}}
(IVOA), the Space-Time Coordinate metadata (STC) provides an alternative to
accurately describing shapes on the celestial sphere. An STC region
has both XML (STC-X) and string (STC-S) serializations that are well mapped
onto our data structures, and translators are being implemented to support them.
The STC-X parser is deployed and being tested as part of the the National Virtual
Observatory's
(NVO%
\footnote{Click to visit \url{http://us-vo.org}})
Footprint Service%
\footnote{Click to visit \url{http://voservices.net/footprint}}
and the STC-S capabilities shall be part of future releases
of our software packages and the service.
The plain string representation STC-S is very straightforward
by default. The union of a set of convexes looks like
\begin{center}\begin{minipage}{8cm}{\footnotesize\begin{verbatim}
Union J2000
(
    Convex   1 0 0   0.1
            -1 0 0  -0.5
    Convex   0 1 0   0.0
)
\end{verbatim}}\end{minipage}\end{center}
The simplicity of the representation comes from the defaults
that are substituted automatically for the missing STC elements
but it is also possible to be explicit about every detail and even
have different coordinate systems for the components, e.g., in
\begin{center}\begin{minipage}{8cm}{\footnotesize\begin{verbatim}
Union
(
    Polygon FK5 J2000  180 0  182 0  182 2  180 2
    Circle  FK4 B1950  179 0  2
)
\end{verbatim}}\end{minipage}\end{center}
In addition to the union, STC sports a large set of features that
also include intersection, difference and negation of any
of the shapes.

The IVOA also working toward a standardized search language
called the Astronomical Data Query Language (ADQL),
which is an IVOA Recommendation as of this writing.
ADQL is an extended subset of the ANSI SQL-92 standard, which adds
geometrical constraints via simplified GIS-like functions (e.g., circle)
and the full support of aforementioned STC representation.
Building on our legacy SQL routines that currently work with the internal string
representation, we are now creating new versions that input STC-S, and are
designing an efficient implemention that supports the full ADQL.

\section{Summmary} \label{sec:sum}

We explored the properties of generalized spherical polygons.
We found the convex representation to be very compact and efficient.
The boolean region operations are straightforward within this framework
but inevitably create convoluted and redundant descriptions.
The simplification involves solving the spherical geometry
of the region. In the process, we summarized how to analytically
calculate the areas and create the outlines of the regions.

Pixelization schemes also significantly benefit from the better
representation. An optimal set of HTM triangles can be computed to
cover entirely a region, or just to approximate the inside, in much less
time than before.
We introduced a heuristic hybrid solutions for simplifications based
on the Igloo pixelization hierarchy that can tackle massive geometries
where the brute-force method fails to deliver.

Our portable implementation in the .NET programming model is available
for developers on the projects website.
Its design enables quick development of new applications and a clean and
easy integration with the SQL Server database engine that hosts several
science archives today. The provided working examples illustrate common
search patterns in C\# and SQL.
Future works include more algorithmic optimizations and full support
of the IVOA's STC and ADQL standards.

\acknowledgements %
The authors are grateful to Arnold Rots for his heroic work on the STC standard,
as well as to Gretchen Greene, Steve Lubow and Rick White for their feedback
on our software packages deployed in the HLA.
The authors acknowledge generous support from the following organizations:
Gordon and Betty Moore Foundation GBMF 554,
W.~M. Keck Foundation KECK D322197,
NSF NVO AST-0122449,
NASA AISRP NNG05GB01G.

\end{document}